\documentclass[12pt]{article}
\usepackage{graphicx}
\usepackage{epsfig}
\usepackage{bm}
\usepackage{amsfonts}
\usepackage{color}
\def\be {\begin{equation}}
\def\ee {\end{equation}}
\def\ba {\begin{eqnarray}}
\def\ea {\end{eqnarray}}

\def\bi {\begin{itemize}}
\def\ei {\end{itemize}}
\begin{document}
\def\bea{\begin{eqnarray}}
\def\eea{\end{eqnarray}}

\title{\textbf{Cardy-Verlinde formula of Kehagias-Sfetsos black hole
}}

\author{  \textbf{M. R. Setare} \thanks{%
E-mail: rezakord@ipm.ir}\\{Department of Science, Payame Noor University, Bijar, Iran} \\
\textbf{Mubasher Jamil} \thanks{%
E-mail: mjamil@camp.nust.edu.pk
 }\\
{ Center for Advanced Mathematics and Physics, National University
of }\\{Sciences and Technology,  Rawalpindi, 46000, Pakistan}}

\maketitle

\begin{abstract}
\textbf{Abstract:} In this paper, we have shown that the entropy of
the Kehagias-Sfetsos black hole in
Ho$\check{\textbf{r}}$ava-Lifshitz (HL) gravity can be expressed by
the Cardy-Verlinde formula. The later is supposed to be an entropy
formula of conformal field theory in any dimension.
\end{abstract}\maketitle
\newpage
\section{Introduction}

Verlinde put forward a very interesting formula \cite{verlinde}
which relates the entropy of conformal field theory in arbitrary
dimension to its total energy and Casimir energy. It has been shown
to hold for topological Reissner-Nordstr\"{o}m \cite{set1} and
topological Kerr-Newman \cite{set2} black holes  in de Sitter
spaces, Taub-Bolt-${AdS}_{4}$ \cite{tab}, Kerr-(A)dS \cite{kle} and
BTZ black hole \cite{jamil}. There are many other relevant papers on
the subject \cite{set3,set4,a1}. Thus, one may naively expect that
the entropy of all CFTs that have an AdS-dual description is given
as the form. However, AdS black holes do not always satisfy the
Cardy-Verlinde formula \cite{gib}. The aim of this paper is to
further investigate the AdS/CFT correspondence in terms of
Cardy-Verlinde entropy
formula.\\
Recently, a power-counting renormalizable, ultra-violet (UV)
complete theory of gravity was proposed by Ho\v{r}ava in
\cite{hor2,hor1,hor3,hor4}. Although presenting an infrared (IR)
fixed point, namely General Relativity, in the  UV the theory
possesses a fixed point with an anisotropic, Lifshitz scaling
between time and space. Due to these novel features, there has been
a large amount of effort in examining and extending the properties
of the theory itself
\cite{Volovik:2009av,Orlando:2009en,Nishioka:2009iq,
Konoplya:2009ig,Charmousis:2009tc,Li:2009bg,Visser:2009fg,Sotiriou:2009gy,
Sotiriou:2009bx,Germani:2009yt,Chen:2009bu,Chen:2009ka,Shu:2009gc,
Bogdanos:2009uj,Kluson:2009rk,Alexandre:2009sy,
Blas:2009qj,Capasso:2009fh,Chen:2009vu,Kluson:2009xx}. Additionally,
application of Ho\v{r}ava-Lifshitz gravity as a cosmological
framework gives rise to Ho\v{r}ava-Lifshitz cosmology, which proves
to lead to interesting behavior
\cite{Calcagni:2009ar,Kiritsis:2009sh, bin}. In particular, one can
examine specific solution subclasses \cite{Lu:2009em,Nastase:2009nk,
Minamitsuji:2009ii,Wu:2009ah,Cho:2009fc,Boehmer:2009yz, SM1}, the
phase-space behavior \cite{Carloni:2009jc,Leon:2009rc}, the
gravitational wave production
\cite{Mukohyama:2009zs,Takahashi:2009wc,Koh:2009cy,Park:2009gf,Park:2009hg},
the perturbation spectrum
\cite{Mukohyama:2009gg,Piao:2009ax,Gao:2009bx,Chen:2009jr,
Gao:2009ht,Kobayashi:2009hh,Wang:2009azb, ding}, the matter bounce
\cite{Brandenberger:2009yt,Brandenberger:2009ic,Suyama:2009vy,
Cai:2009in}, the black hole properties
\cite{Danielsson:2009gi,Cai:2009pe,Kehagias:2009is,Mann:2009yx,
Bertoldi:2009vn,Park:2009zra,Castillo:2009ci,BottaCantcheff:2009mp,
Varghese:2009xm,Kiritsis:2009rx, mhi}, the dark energy phenomenology
\cite{Saridakis:2009bv,Appignani:2009dy,Setare:2009vm}, the
astrophysical phenomenology
\cite{Kim:2009dq,Harko:2009qr,Iorio:2009ek}, the thermodynamic
properties \cite{Wang:2009rw,Cai:2009qs} etc. However, despite this
extended research, there are still many ambiguities if
Ho\v{r}ava-Lifshitz gravity is reliable and capable of a successful
description of the gravitational background of our world, as well as
of the cosmological behavior of the universe
\cite{Charmousis:2009tc,Li:2009bg,Sotiriou:2009bx,Bogdanos:2009uj,Koyama:2009hc}.
In the present paper we would like to check the consistency of the
Cardy-Verlinde formula, for the Kehagias-Sfetsos black hole.
\section{Kehagias-Sfetsos black hole}
The natural setting of Ho\v{r}ova-Lifshitz gravity is the ADM
formalism, where the four dimensional metric is parameterized by the
following
\begin{equation}
ds^2=-N^2c^2dt^2+g_{ij}(dx^i+N^idt)(dx^j+N^jdt).
\end{equation}
Here $N$ is the lapse function and $N_i$ is the shift function,
respectively. The Ho\v{r}ova action is
\begin{eqnarray}
S&=&\int dt dx^3\sqrt{g}N\Big[
\frac{2}{\kappa^2}(K_{ij}K^{ij}-\lambda_gK^2)
-\frac{\kappa^2}{2\nu^4_g}C_{ij}C^{ij}
+\frac{\kappa^2\mu}{2\nu^4_g}\epsilon^{ijk}
R_{il}^{(3)}\nabla_jR^{(3)l}_k\nonumber\\&&
-\frac{\kappa^2\mu^2}{8}R_{ij}^{(3)}R^{(3)ij} +
\frac{\kappa^2\mu^2}{8(3\lambda_g-1)} \Big(
\frac{4\lambda_g-1}{4}(R^{(3)})^2-\Lambda_WR^{(3)}+3\Lambda_W^2\Big)\nonumber\\&&
+\frac{\kappa^2\mu^2 w}{8(3\lambda_g-1)}R^{(3)}\Big],
\end{eqnarray}
where $\kappa$, $\lambda_g$, $\nu_g$, $\mu$, $w$ and $\Lambda_W$ are
constant parameters. Also $R^{(3)}$ is a 3-dimensional curvature
scalar for $g_{ij}$; $K_{ij}$ is the extrinsic curvature given by
\begin{eqnarray}
K_{ij} = \frac{1}{2N} \left( {\dot{g_{ij}}} - \nabla_i N_j -
\nabla_j N_i \right) \, ,
\end{eqnarray}
is the extrinsic curvature and
\begin{eqnarray}
C^{ij} \, = \, \frac{\epsilon^{ijk}}{\sqrt{g}} \nabla_k \bigl( R^j_i
- \frac{1}{4} R \delta^j_i \bigr),
\end{eqnarray}
is the Cotton-York tensor. The fundamental constants including speed
of light $c$, Newton's gravitational constant $G$ and the
cosmological constant $\Lambda$ are defined as
\begin{equation}
c^2=\frac{\kappa^2\mu^2|\Lambda_W|}{8(3\lambda_g-1)^2},\ \
G=\frac{\kappa^2c^2}{16\pi(3\lambda_g-1)},\ \
\Lambda=\frac{3}{2}\Lambda_Wc^2.
\end{equation}
Consider a static and spherically symmetric solution given by
\begin{equation}
ds^2=-e^{\nu(r)}dt^2+e^{\lambda(r)}dr^2+r^2(d\theta^2+\sin^2d\phi^2),
\end{equation}
By substituting the above metric ansatz into the action, the
resulting reduce Lagrangian is given by
\begin{eqnarray}
\mathbb{L}&=&\frac{\kappa^2\mu^2}{8(1-3\lambda_g)}e^{(\nu+\lambda)/2}\Big[
(2\lambda_g-1)\frac{(e^{-\lambda}-1)^2}{r^2}+2\lambda_g\frac{e^{-\lambda}-1}{r}
(\lambda'e^{-\lambda})+\frac{\lambda_g-1}{2}(\lambda'e^{-\lambda})^2\nonumber\\&&
-2(w-\Lambda_W)(1-e^{-\lambda}(1-r\lambda'))-3\Lambda^2_Wr^2
\Big].
\end{eqnarray}
The above Lagrangian yields the following equations of motion:
\begin{eqnarray}
0&=&(2\lambda_g-1)\frac{(e^{-\lambda}-1)^2}{r^2}-2\lambda_g\frac{e^{-\lambda}-1}{r}
(\lambda'e^{-\lambda})+\frac{\lambda_g-1}{2}(\lambda'e^{-\lambda})^2\nonumber\\&&
-2(w-\Lambda_W)[1-e^{-\lambda}(1-r\lambda')]-3\Lambda^2_Wr^2,
\end{eqnarray}
\begin{eqnarray}
0&=&\frac{\nu'+\lambda'}{2}\Big[
(\lambda_g-1)(\lambda'e^{-\lambda})-2\lambda_g\frac{e^{-\lambda}-1}{r}+2(w-\Lambda_W)r
\Big]\nonumber\\&&+(\lambda_g-1)\Big[
(-\lambda''+\lambda')e^{-\lambda}-2\frac{e^{-\lambda}-1}{r^2} \Big],
\end{eqnarray}
by varying functions $\nu$ and $\lambda$ respectively.\\
Now by imposing $\lambda_g=1$, which reduces to the Einstein-Hilbert
action in the infra-red limit, one obtains the following solution of
the vacuum field equations in Ho\v{r}ava gravity:
\begin{equation}
e^{\nu(r)}=e^{-\lambda(r)}=1+(w-\Lambda_W)r^2-\sqrt{r[w(w-2\Lambda_W)r^3+\beta]}.
\end{equation}
Here $\beta$ is an integration constant. Now the Kehagias-Sfetsos
(KS) black hole solution \cite{Kehagias} is obtained by considering
$\beta=4wM$ and $\Lambda_W=0$,
\begin{equation}
e^{\nu(r)}=1+wr^2-wr^2\sqrt{1+\frac{4M}{wr^3}}.
\end{equation}
If we impose the limit $\frac{4M}{wr^3}\ll1$, then the last
expression yields the Schwarzschild metric $e^{\nu(r)}=1-2M/r$.
There are two horizons
\begin{equation}
r_\pm=M\Big[1\pm\sqrt{1-\frac{1}{2wM^2}}\Big].
\end{equation}
To avoid a naked singularity at the origin, impose the condition
$wM^2\gg1$, the outer horizon approaches the Schwarzschild horizon
$r_+\simeq2M$, and the inner horizon approaches the central
singularity, $r_-\simeq0$.

\section{Cardy-Verlinde formula}

In this section, we introduce the Cardy-Verlinde formula which
states that the entropy of a (1+1)-dimensional CFT is given by
\begin{equation}
S=2\pi\sqrt{\frac{c'}{6}\Big( L_0-\frac{c'}{24}  \Big)},
\end{equation}
where $c'$ is the central charge and $L_0$ is the Virasoro
generator. After appropriate identifications of $c'$ and $L_0$, the
above Cardy formula, we obtain the generalized Cardy-Verlinde
formula which takes the form \cite{verlinde}
\begin{equation}
S_{CFT}=\frac{2\pi R}{\sqrt{ab}}\sqrt{E_C(2E-E_C)},
\end{equation}
where $E$ is the total energy, $E_C$ is the Casimir energy, $a$ and
$b$ are arbitrary positive constants. Also $R$ is the radius of the
$n+1$ dimensional spacetime, $ds^2=-dt^2+R^2d\Omega_n$.
 The definition of
Casimir energy is derived by the violation of the Euler relation as
\begin{equation}
E_C=n(E+PV-TS-\Phi Q-\Omega J),
\end{equation}
where the pressure of the CFT is given by $P=E/nV$. For KS black
hole, $J$ and $Q$ are zero. The total energy is the sum of two terms
\begin{equation}
E(S,V)= E_E(S,V)+\frac{1}{2} E_C(S,V).
\end{equation}
Here $E_E$ is the purely extensive part of the total energy. The
Casimir energy and the purely extensive part of the total energy are
expressed as
\begin{equation}
E_C=\frac{b}{2\pi R}S^{1-\frac{1}{n}},
\end{equation}
\begin{equation}
E_E=\frac{a}{4\pi R}S^{1+\frac{1}{n}}.
\end{equation}

\section{Entropy of Kehagias-Sfetsos black hole and Cardy-Verlinde formula}

Following \cite{cao}, the entropy of the KS black hole is assumed to
be given by $S=A/4=\pi r_+^2$, which yields
\begin{equation}
S=\pi M^2\left[1+\sqrt{1-\frac{1}{2wM^2}}\right]^2.
\end{equation}
Padmanabhan \cite{paddy} has shown that temperature of the event
horizon of a spherically symmetric spacetime is given by
\begin{eqnarray}
T&=&\frac{1}{4\pi}\frac{\partial e^{\nu(r)}}{\partial
r},\nonumber\\&=&\frac{w}{2\pi}\left[
r_+-r_+\sqrt{1+\frac{4M}{wr_+^3}}+\frac{3M}{r_+^2\sqrt{1+\frac{4M}{wr_+^3}}}
\right],
\end{eqnarray}
calculated at the event horizon $r=r_+.$
 We choose $n=2$ and
$E=M$ for KS black hole. The Casimir energy becomes
\begin{eqnarray}
E_C&=&3M-2TS,\nonumber\\
&=&3M-wM^2\left[
r_+-r_+\sqrt{1+\frac{4M}{wr_+^3}}+\frac{3M}{r_+^2\sqrt{1+\frac{4M}{wr_+^3}}}
\right]\left[1+\sqrt{1-\frac{1}{2wM^2}}\right]^2.
\end{eqnarray}
The pure extensive part of the total energy is given by
\begin{eqnarray}
E_E&=&-\frac{1}{2}M+TS,\nonumber\\
&=&-\frac{1}{2}M+\frac{wM^2}{2}\left[
r_+-r_+\sqrt{1+\frac{4M}{wr_+^3}}+\frac{3M}{r_+^2\sqrt{1+\frac{4M}{wr_+^3}}}
\right]\left[1+\sqrt{1-\frac{1}{2wM^2}}\right]^2.\nonumber\\
\end{eqnarray}
Also
\begin{eqnarray}
2E-E_C&=&-M+2TS,\nonumber\\
&=&-M+wM^2\left[
r_+-r_+\sqrt{1+\frac{4M}{wr_+^3}}+\frac{3M}{r_+^2\sqrt{1+\frac{4M}{wr_+^3}}}
\right]\left[1+\sqrt{1-\frac{1}{2wM^2}}\right]^2.\nonumber\\
\end{eqnarray}
From comparison of equations (17) and (21), we obtain
\begin{eqnarray}
R&=&\frac{bS^{1/2}}{4\pi}\Big( \frac{3}{2}M-TS  \Big)^{-1},\nonumber\\
&=&\frac{b}{4\pi^{1/2}} M\Big(1+\sqrt{1-\frac{1}{2wM^2}}\Big)\Big[
\frac{3}{2}M-\frac{wM^2}{2}\Big(
r_+-r_+\sqrt{1+\frac{4M}{wr_+^3}}+\frac{3M}{r_+^2\sqrt{1+\frac{4M}{wr_+^3}}}
\Big)\nonumber\\&&\times\Big(1+\sqrt{1-\frac{1}{2wM^2}}\Big)^2\Big]^{-1}.
\end{eqnarray}
From comparison of equations (18) and (22), we obtain
\begin{eqnarray}
R&=&\frac{aS^{3/2}}{4\pi}\Big( -\frac{1}{2}M+TS \Big)^{-1},\nonumber\\
&=&\frac{a}{4} \pi^{1/2}
M^3\Big(1+\sqrt{1-\frac{1}{2wM^2}}\Big)^3\Big[-\frac{1}{2}M+\frac{wM^2}{2}\Big(
r_+-r_+\sqrt{1+\frac{4M}{wr_+^3}}+\frac{3M}{r_+^2\sqrt{1+\frac{4M}{wr_+^3}}}
\Big)\nonumber\\&&\times\Big(1+\sqrt{1-\frac{1}{2wM^2}}\Big)^2
\Big]^{-1}.
\end{eqnarray}
Taking product of (24) and (25), we obtain
\begin{eqnarray}
R&=&\frac{\sqrt{ab}}{4\pi}\frac{S}{\sqrt{\Big( \frac{3}{2}M-TS
\Big)^{-1}\Big( -\frac{1}{2}M+TS \Big)^{-1}}}.
\end{eqnarray}
Using (21), (22) and (26) in (14), we obtain
\begin{equation}
S_{CFT}=S.
\end{equation}
It has been proven that the entropy of the Kehagias-Sfetsos black
hole can be expressed in the form of Cardy-Verlinde formula. We
would like to remark that the technique used above to prove this
result has been followed earlier for the rotating charged BTZ black
hole as well \cite{jamil,set4}. Moreover, the quantity $R$ is in
general arbitrary but in the presence of a background geometry, it
becomes particular and therefore can be written in terms of the
parameters of Kehagias-Sfetsos spacetime (the geometry), like we
obtained in (26).

\section{Conclusion}

HL theory brings some important new features from the GR to the
higher dimensional lagrangian and it's role in construction a non
relativistic candidate for quantum gravity. According to the Blas
et al arguments \cite{Blas:2009qj}, it seems that this model must
be modified by some terms to avoiding from strong coupling,
instabilities, dynamical in consistencies and unphysical extra
mode. One of the first exact solutions for this modified version
is the work of Kiritsis \cite{8}. Indeed the Kiritsis work
contains some previous families of exact solutions as a special
sub class and has a good asymptotic behaviors. The explicit form
of exact solution for this modified version deal with some
algebraic quadratures and lead finally to an implicit static
spherically symmetric metric. But no doubt this solution generic,
avoids from the trouble problems which occur in the original
version of HL. As we know that there are 2 explicit family of
exact solutions for a spherically symmetric background without
projectability condition and other solutions all are the familiar
GR solution i.e $Ads_{4}$-Schwarzchild solutions. One solution
belongs to the \cite{Kehagias} which in abberation in literatures
known as KS solution. This solution is asymptotically flat and as
we showed that in spite of the GR BHs, it's timelike geodesics is
stable \cite{sm}.
 The aim of this paper
is to further investigate the AdS/CFT correspondence in terms of
Cardy-Verlinde entropy formula. We have shown that the entropy of
the black hole horizon of Kehagias-Sfetsos spacetime can also be
written in the form of Cardy-Verlinde entropy formula.

\end{document}